# Enabling Scientific Crowds: The Theory of Enablers for Crowd-Based Scientific Investigation

Jorge M. Faleiro Jr. #1

#Centre of Computational Finance and Economic Agents, University of Essex
Wivenhoe Park, Colchester, CO4 3SQ, UK
1 jfalei@essex.ac.uk   j@falei.ro

*Abstract[1]*— Evidence shows that in a significant number of cases the current methods of research do not allow for reproducible and falsifiable procedures of scientific investigation. As a consequence, the majority of critical decisions at all levels, from personal investment choices to overreaching global policies, rely on some variation of try-and-error and are mostly non-scientific by definition. We lack transparency for procedures and evidence, proper explanation of market events, predictability on effects, or identification of causes. There is no clear demarcation of what is inherently scientific, and as a consequence, the line between fake and genuine is blurred.

This paper presents highlights of the Theory of Enablers for Crowd-Based Scientific Investigation, or Theory of Enablers for short. The Theory of Enablers assumes the use of a next-generation investigative approach leveraging forces of human diversity, micro-specialized crowds, and proper computer-assisted control methods associated with accessibility, reproducibility, communication, and collaboration.

This paper defines the set of very specific cognitive and non-cognitive *enablers* for crowd-based scientific investigation: *methods of proof*, *large-scale collaboration*, and a domain-specific *computational representation*. These enablers allow the application of procedures of structured scientific investigation powered by crowds, a collective brain in which neurons are human collaborators

## I. BACKGROUND AND ASSUMPTIONS

Economics and finance are particular domains of knowledge in which related systems and agents – markets, humans and their relationships – are hard, if not impossible, to model [1]. For the appropriate investigation, adequate financial models must be able to deal with this intrinsic complexity of economic systems and agents [2] [3] [4].

This research advocates the use of crowds for the investigation and resolution of complex problems in general and in economics in particular, an approach we are calling *crowd-based investigation* [5]. At the point of this writing, available literature suggests that the suitability of crowds for the resolution of complex problems can only be confirmed by empirical evidence [5]. We consider that the definition of mechanisms that should be in place to allow the use of crowds for the resolution of complex problems to be mostly axiomatic, based primarily on three specific assumptions:

- The process by which we acquire objective knowledge must follow the rules dictated by the modern scientific method. As a consequence, the proof of observations as being real or false must be driven by a widely known set of quantifiable standards [5].

- Human collaboration in large-scale is an adequate method to investigate and resolve complex problems, and collaboration in large-scale is enabled by providing the correct set of incentives to crowd participants [5].

- Computers should fulfill the role of a tool to support discovery and should not serve as a replacement for the application of reproducible and falsifiable procedures of the scientific method.

## II. ENABLERS

The specific assumptions listed in Section I bring two immediate consequences concerning a method for resolution of complex problems in general:

- The need for an investigation method that applies to large groups of individuals, or crowds, for the resolution of complex problems, or problems of difficult representation through models. We are calling methods of investigation that apply to collaborative investigation methods of *crowd-based investigation*.

- The scientific effectiveness of an investigation based on crowds is related to the existence of specific environmental requisites that must be in place to allow, but not necessarily guarantee, the application of a proper scientific method by crowds of individuals[2].

We are calling these enabling requisites *enablers of a crowd-based investigation*.

Enablers are classified as either cognitive or non-cognitive. The association between a crowd-based method of investigation and enablers is shown in Figure 1.

---

[1] Large portions of this paper are reproduced as part of [1] [5]

[2] The implicational relationship of the statement is that the existence of enablers in an environment is a necessary but not sufficient condition for the proper support of crowd-based investigation.

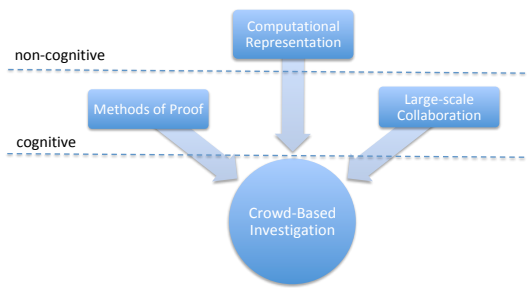

**Figure 1. Enablers of Crowd-Based Methods of Scientific Investigation**

Cognitive and non-cognitive requisites, or enablers, of the next wave of investigation methods based on crowds: methods of proof; large-scale collaboration; and a computational representation.

*Cognitive enablers* relate to non-computational features associated with the subjective mechanisms of human understanding of what to consider knowledge and the underlying fabrics of large-scale collaboration. Cognitive enablers are not domain specific, and as a consequence should be the same regardless of the domain of knowledge under consideration [1]. Cognitive enablers are *methods of proof* and *large-scale collaboration* [5].

*Non-cognitive enablers*, on the other hand, relate to features that can be directly and purely mapped to a computational description. This description is called a *computational representation* [6]. Unlike cognitive enablers, non-cognitive enablers are domain specific, and as a consequence, each domain of knowledge must be supported by a different, specially tailored, computational representation [7] [8].

### III. METHODS OF PROOF

Humans learn new things through investigation. Careful investigation is what establishes if an observed phenomenon is real, or it should be deemed just a result of random forces of nature at play. The primary target of any investigation is to establish facts, as accurately as possible, by proving observations to be either true or false. That is how humankind has been accumulating objective knowledge for as long as we walk this earth, and this is why defining precise methods of proof is crucial.

The mental process we follow as individuals to investigate and learn about things is not straightforward. Even at this present date, science is still not able to unequivocally explain the process by which we learn and assess things. If this is true when we produce our thoughts on our own, we should expect an even more elaborate process to be at play when we introduce procedures of investigation that are performed by multiple individuals, organized in seemingly chaotic crowds [5].

Given the number of participants and the nature of the interaction – formal scientific investigation - we can safely expect as a consequence a large number of hypotheses being generated and tested. On this scenario, ideas must be defined, exchanged, discussed, and tested in a sequence of steps, arranged like a pipeline [5]. Procedures in each step of the pipeline should potentially generate massive amounts of data, and each piece of data should be unquestionably tested as true or false. As a consequence, each of the steps must abide by standards and validation metrics that must be well understood and accepted by all participants [5].

A simplistic description of such a pipeline would be a tube, where its input, taken on the head of the pipe, is a problem, or a set of ideas under investigation, and other intangible aspects such as the experience of the individual performing the inquiry or the investigation. On the tail of the tube, the result of the investigation, as either true or false. Over the extension of the tube, there are small holes, from where the process produces pre-defined, controlled evidence [5].

The idea of arranging a sequence of pre-defined steps to assert a result of an investigation as true or false is not new. There are references in the literature to a step-by-step process in biomedical research, specifically for statistical measurements, referred to as a "statistical pipeline" [9] [10]. Although similar in its overreaching purpose and the intended standardization of the understanding of what is true or false, the scope of what that pipeline would entail is different than what this research proposes. Their scope is also limited specifically to software patterns and a computational platform. In the field of economics, there are proposals in the literature with a minor overlapping with the idea of proof pipelines, arranging economic models as testable pieces of engineering, not necessarily as pipelines, referred to as "economic wind tunnels" [11].

### IV. LARGE-SCALE COLLABORATION

The use of crowds for resolution of problems follows one of two distinct approaches. The first approach, named "wise crowds" [12] relies on empirical observations [13] [14] and assumes the existence of some invisible, unquantifiable mechanism, somehow providing a certain level of knowledge to crowds, therefore allowing them to make wise decisions. The "wise crowd" approach relies on the assumption of complete independence and decentralization between participants of a crowd. The second approach, named *collaborative crowds*, assumes that knowledge is produced as a result of structured collaboration between participants of a crowd.

The theory of enablers assumes a second approach, collaborative crowds, where large-scale collaboration occurs by the existence of particular requirements of collaboration, as a natural evolutionary response to the environment in which investigation takes place.

The use of crowds as agents of investigation requires an organization of a large number of individuals, in different roles and at different levels of technical understanding, to continually collaborate for the resolution of complex problems. However, as we can readily ascertain by observation, collaboration does not come out of thin air. We need something to drive effective collaboration, and in this section, we concentrate on explaining requirements for effective collaboration to take place.

Collaboration is what builds "some sort of a collective brain with the people in the group playing the role of neurons" [15] [12] and ultimately amplifies the intelligence of a group of people. Collaboration is facilitated as a result of four *requirements*: expert attention, proper cultural and intellectual development, manufactured serendipity, and human diversity.

One of the assumptions of the theory of enablers, listed in Section I, is that organized human collaboration is well suited for the investigation and resolution of complex problems. In reality, it would be impossible to infer absolute suitability of large-scale collaboration for the resolution of complex problems. Alternatively, we can enumerate results from empirical exercises, and their specific details, as evidence of resolution of complex problems by crowds [5].

## V. Computational Representation

Similarly to natural languages, a computational representation grows from the needs of a specialized domain of knowledge and therefore is suited for use cases relevant to that specific domain [7] [8] [16].

In some domains of knowledge, like architectural sciences, one would be more concerned about spaces, shapes, volumes or colors, and their relationships with a three-dimensional environment and the effect of the interaction of those concepts with humans. In legal sciences, one would be more concerned about possible associations between real-world entities, and rules defining their behavior and constraints for interaction. In some other domains, like bioinformatics, the ability to represent interconnected shapes and strings could be more relevant. In biophysics, it is essential to keep track of genotypical and phenotypical traits, and their relationships with encoded protein sequences with a vast number of possible combinations. In economics, our subject of concern, a researcher would be more interested in the way changes in quantitative measurements, over a time series, would affect the valuation [7].

A computational representation must mimic the inherently free flow of thoughts of the human mind and the speed of modern vehicles of collaboration, and therefore, by similarity, a computational representation must be fluid. In contradiction, computational artifacts, like programming languages and databases, are born out of strictly technical aspects of a problem and bred outside of concerns relevant to specific domains of knowledge. Only after definition, they are forcedly introduced for use and therefore not able to follow the free-flow of the evolution of ideas. Computational artifacts remain frozen to domain-specific requirements of that specific point in time when the introduction occurred. When requirements on that domain evolve to follow the increasing complexity of the problems at hand, those artifacts would no longer fit, or in a best case require an additional verbosity, sacrificing the proper semantics of communication [7].

In opposition to computational artifacts, a computational representation must be dynamic, able to adapt and evolve to solve new classes of problems and organize increasingly complex and powerful computing environments. These new classes of problems are different from the problems we had to deal with just a few years back. They require the collaboration of multi-disciplinary specialists exchanging different types of artifacts that must be adequately described and tracked [16]. An investigator must have adequate tools and methods to approach new problems correctly. On this sense, an adequate computational representation allows for the proper description and control of those tools and methods, allowing them to change in the face of new demands and be able to address new problems [8].

## VI. Future Work

The theory of enablers offers an approach for large-scale crowd collaboration to potentially circumvent the inherent problems related to the information crisis we currently observe in the scientific investigation [16]. These foundations are a significant improvement over the current methods of investigation in finance and offer opportunities for further research on the same subject are far from exhausted.

The subject of modern methods of investigation relies mostly on mechanisms of extreme complexity, and an approach to find for a resolution of a complex problem is to search a prospective solution through organic and incremental iterations [17] [18] [19]. This incremental, iterative approach is based on the classical principle of "design thinking" [20]. The main concerns of iterations of future research would be related to:

- A working minimum collaborative environment for crowd-based investigation;
- Versioning of complex run-time graphs;
- Quantification of features of collaboration.

With that, the initial step in an iterative approach would be the definition of a minimum environment in which the original ideas of this research could be exposed to an initially restricted group of participants and tested. A platform to support a controlled interaction, even if for a small and controlled number of research participants, will help explore algorithms and methods to record contributions and their evidential properties.

The use of evidential properties would require, amongst other things, the support of a record of provenance [16] [8] [1]. The record of provenance would require storing and versioning of the representation of the financial model as a graph associated with the stream of the execution [1].

Versioning of complex graphs by a simple full copy of the entire graph on change of any attributes of the graph will trigger a space explosion. Some alternatives have been recently proposed for a particular type of graph called property graphs [21]. Property graphs are oversimplified, compared to a graph that intends to represent a flow of execution that must be materialized on different execution spaces, and as a consequence can be materialized on distinct, distributed processors [7]. An alternative to in-place versioning would be to apply changes to graphs by separate clone and merge operations, in a distributed approach [22]. The drawback of a clone-and-merge approach is that, since processors and endpoints are in nature pervasive[3], a clone of a single graph can force cloning of others. The feasibility of a clone and a subsequent merge operation is limited by how large the affected graph might be.

For example, if both financial models $\phi_1$ and $\phi_2$ use a processor $P$, then a clone of $\phi_1$ will force clone of $\phi_2$, and the same for the merge operation. Due to other processors that can be possibly involved, the cloning chain can potentially involve other graphs. As a consequence, for real-world investigations, the final full size of the final chain can become an impediment for clone and merge operations.

Even if limited, in case not all of these questions are finally and adequately addressed, a minimum platform allowing a rudimentary interaction amongst participants of a scientific investigation is still useful for the investigation and additional models and allows two immediate consequences. First, since a minimum platform would allow for reproducibility and traceability of evidence through the use of contributions, it should serve as an incentive for research to be conducted on the platform. Second, new financial models, or cases of use, would show the need for additional facets and possibly even contributions, extending the current computational representation.

The use of crowds for resolution of problems follows one of two distinct approaches. The first approach, named "wise crowds" [12] relies on empirical observations [13] [14] and assumes the existence of some invisible, unquantifiable mechanism, somehow providing a certain level of knowledge to crowds, therefore allowing them to make wise decisions. The "wise crowd" approach relies on the assumption of complete independence and decentralization between participants of a crowd. The second approach, named *collaborative crowds*, assumes that knowledge is produced as a result of structured collaboration between participants of a crowd.

This research subscribes to the second approach, collaborative crowds, where collaboration in large scale occurs by the existence of particular requirements of collaboration [5]. It is necessary to produce metrics and quantify the requirements that must be in place for large-scale collaboration, but proper quantification can only happen in a real collaboration environment, where proper participants are engaged. The quantification of requirements for large-scale collaboration, what we call *collaboration metrics*, would allow to measure the potential efficacy of a disjoint group of researchers and compare the performance of different investigation exercises on items that are specifically related to how well collaboration takes place.

Quantifiable requirements would affect one or more *features of collaboration*. Features of collaboration would give, as a group, indications on the efficiency of the scientific investigation performed by the crowd, for a given specific investigation. Not all requirements for collaboration are subject to quantification [5]. Proposed quantifiable requirements for future research are the level of micro-expertise attention and cognitive diversity.

The first quantifiable requirement, the level of micro-expertise attention, is crucial to collaboration, and as a consequence, also to collaborative crowds[4]. The level of micro-expertise attention would measure specific features of collaboration: the relevancy of a participant in a crowd, per investigation subject; the quality of contributions produced per participant; and the influence of a participant.

Consider for illustration purposes a directed graph $\phi = (C, E)$ of vertices $C$ and edges $E$. In $\phi = (C, E)$, $C$ is a set of contributions $(c_0, \ldots, c_n)$, and $E$ is a set of edges $(e_0, \ldots, e_n)$, indicating dependencies between contributions. Since contributions are produced as a result of a financial model[5], a contribution $c'$ is considered a dependency of $c''$ if $c''$ is upstream to $c'$ (in the sense that a financial model is a stream). A specific financial model might be tagged as belonging to zero or more subjects of investigation. Given the streaming nature of a financial model, this determination is straightforward. Features of collaboration are taken from the application of specific graph algorithms to the graph $\phi = (C, E)$:

- The in-degree centrality[6] of the contribution $c_i$ would quantify the quality of that specific contribution;
- The relevancy of a participant in a crowd, per subject of investigation, is given by the summation, of the quality of all contributions $C$ created by that participant;
- The level of influence of a participant is the summation of the relevancy of that participant, across all subjects of investigation.

The second quantifiable requirement is cognitive diversity. The literature points to metrics for cognitive diversity based on cognitive distance [23] [24] [25]. Future research should also assume a possible reliance on the interaction of participants, or "texts of utterances of a collective's member" [23], for measurement of a crowd's cognitive diversity. Another possibility, given recent literature just published, would be the use of a more sophisticated model of sentiment and emotion analysis using unstructured data [26]. This research could not find in the literature a proposal for quantification of cognitive diversity for scientific investigation.

This list of future research opportunities is not final. The availability of a platform for crowd-based investigation would allow the collection of evidence to provide new insights into metrics and algorithms for additional items of research.

---

[3] Given the intent of reusability the same processors and endpoints should potentially repeat themselves in a large number of financial models [7] [1].

[4] "Expert attention is to creative problem solving what water is to life: it's the fundamental scarce resource" [15]

[5] A financial model is also a directed graph [7] [1]

[6] The number of inbound edges to a vertex in a directed graph

## VII. Conclusion

Based on assumptions introduced in Section I this paper defines cognitive and non-cognitive requirements, or enablers, for crowd-based scientific investigation. The enablers for crowd-based investigation are methods of proof, collaboration in large-scale, and a computational representation [1].

The first cognitive enabler, *methods of proof*, is defined in Section III. Methods of proof are required to provide a common framework for a step-wise, algorithmic investigation and standard control points based on statistical methods that apply to a crowd-based investigation. The framework is defined in previous research papers, around something we call a *proof pipeline*, proper methods of proof, the definition of the meaning of scientific learning in economics, and requirements for a specific class of systems called scientific support systems [16] [5] [1].

The second cognitive enabler, *large-scale collaboration*, is defined in Section IV. Large-scale collaboration establishes the mechanics and incentives that must be in place to leverage collaborative crowds. The mechanics and incentives for collaboration are listed in details in previous work. Details include analysis for the suitability of collaborative crowds for the resolution of complex problems and a study of the historical and evolutionary perspective to determine the use of collaborative crowds as the foundation for upcoming methods of scientific investigation [5] [1].

The third non-cognitive enabler, *computational representation*, is defined in Section V. A computational representation is a crowd-friendly representation system to define concepts related to a specific domain of knowledge, and is detailed extensively in previous research work. A computational representation is defined based on facets, contributions, and constraints of data. Facets are definable aspects that make up a subject or an object of a domain of knowledge. Contributions are shareable and formal evidence of a crowd-based scientific investigation. Constraints of data define entities and rules of associations between entities in a specific domain of knowledge. Computational representations are defined based on the steps and procedures outlined in a generic process called a *representational process* [7] [1].

This theory of enablers is primarily about one approach in many in how to improve the way we search for objective knowledge. Knowledge, as it seems, is the result of a recipe by which we mix the contents of a number of different buckets into where we keep distinct perceptions of the world around us: experiences, creative thoughts, hunches, beliefs, biases, and the systematic training. All in some measure is important, and they all have a role in producing objective knowledge. However, the exact recipe, or mechanisms, we follow in mixing the contents of those buckets hides deep in the way our collective minds work and is still an incognita. For that mysterious recipe, we advocate for the use of crowds backed by controlled computational resources to tackle complexity and perform investigations according to the proven guidelines of the scientific method.